\documentclass[%
 reprint,
showpacs,
 amsmath,amssymb,
 aps,prl
]{revtex4-1}

\usepackage{graphicx}
\usepackage{epsfig}
\usepackage{dcolumn}
\usepackage{bm}
\usepackage{newfloat}
\usepackage{amsmath}
\usepackage{amssymb}
\usepackage{amsthm}
\usepackage{textcomp}
\DeclareFloatingEnvironment[name={S}]{suppfig}

\begin{document}


\title{Dissipation in ultrahigh quality factor SiN membrane resonators}
\author{S. Chakram, Y. S. Patil, L. Chang and M. Vengalattore}
 \affiliation{Laboratory of Atomic and Solid State Physics, Cornell University, Ithaca, NY 14853}
\email{mukundv@cornell.edu}

\date{\today}

\begin{abstract}
We study the optomechanical properties of stoichiometric SiN resonators through a combination of spectroscopic and interferometric imaging techniques. At room temperature, we demonstrate ultrahigh quality factors of  $5 \times 10^7$ and a $f \times Q$ product of $1 \times 10^{14}$ Hz that, to our knowledge, correspond to the largest values yet reported for mesoscopic flexural resonators. Through a comprehensive study of the limiting dissipation mechanisms as a function of resonator and substrate geometry, we identify radiation loss through the supporting substrate as the dominant loss process. In addition to pointing the way towards higher quality factors through optimized substrate designs, our work realizes an enabling platform for the observation and control of quantum behavior in a macroscopic mechanical system coupled to a room temperature bath. 
\end{abstract}


\maketitle

Mesoscopic mechanical resonators with ultrahigh quality factors are ubiquitous ingredients in diverse applications of sensing, inertial navigation and communications \cite{li2007,poot2012} as well as in foundational tests of quantum mechanics in macroscopic systems \cite{schwab2005,aspelmeyer2008,kippenberg2008,aspelmeyer2013}. The quantum coherent control of these resonators and the realization of quantum-limited sensors require the cooling of these resonators to low phonon occupancies. However, to date, existing schemes of optomechanical cooling place stringent constraints on the mechanical and optical properties of these resonators. Crucially, overcoming the thermal coupling to the environment and the attainment of long coherence times require low mechanical resonance frequencies, a large `frequency - quality factor product' ($f \times Q > k_B T_{ambient}/h$) and low optical absorption \cite{marquadt2007, wilsonrae2007}. This combination has hitherto been difficult to achieve despite the exploration of a wide range of micro- and nanoscale systems. Thus, recent demonstrations of optomechanical cooling to the mechanical ground state \cite{teufel2011, chan2011} require cryogenic cooling of the mechanical system to reduce the thermal coupling to the environment. 

Stoichiometric silicon nitride membrane resonators show great promise for optomechanics due to their high quality factors and low optical absorption \cite{verbridge2006, thompson2008}. Recent work has demonstrated $f \times Q$ products at room temperature as high as $2 \times 10^{13}$ Hz in these resonators \cite{wilson2009}. However, the dominant loss mechanisms and potential routes to further enhance the quality factors of these resonators remain poorly understood. 

In this Letter, we demonstrate membrane resonators of stoichiometric Silicon nitride with quality factors in excess of $5 \times 10^7$ and a frequency-Q product $f \times Q \sim 1 \times 10^{14}$ Hz that is more than an order of magnitude greater than the requirement for ground state cooling and coherent quantum control of a room temperature optomechanical system. Further, we identify radiation loss through the supporting substrate as the dominant loss mechanism. This finding points to further enhancements of the performance of such membrane resonators through appropriate material choice and design of the supporting substrate. In addition, our resonators are a promising platform for the observation and control of quantum behavior in a mesoscopic mechanical system at room temperature. 

The mechanical oscillators in our study are fabricated by NORCADA Inc., and consist of LPCVD Silicon nitride square membranes under high tensile stress of around 0.8 - 0.9 GPa. The membranes range in thickness from $h \sim 30 - 200 $ nm with lateral dimensions in the range $L \sim 0.5 - 5$ mm. The membranes are deposited on single crystal Silicon wafers. 
The membranes constitute one arm of a Michelson interferometer while the other (reference) arm is actively stabilized against ambient vibrations. This realizes a precise measurement of the instantaneous position of the membrane with a sensitivity of 0.1 pm/Hz$^{1/2}$ for typical powers of 200 $\mu$W incident on the membrane. The optical measurements are performed with a external-cavity diode laser (ECDL) operating at a wavelength of 795 nm. At this wavelength, real  and imaginary parts of the refractive index of the membranes are measured to be $\mathcal{R}(n) = 1.95 \pm 0.02$ and $\mathcal{I}(n) < 10^{-5}$ respectively, leading to a peak reflectivity of 0.34 for the 100nm thick resonators (see Supplementary Information for details). 

From thin plate theory, the mechanical eigenfrequencies of the membranes are $\omega_{jk} = 2 \pi \sqrt{\sigma/4 \rho L^2} \sqrt{j^2 + k^2}$ where $\sigma$ is the intrinsic tensile stress, $\rho = 2.7$ g/cm$^3$ is the mass density and $L$ the lateral dimension of the membrane. We have confirmed this relation is accurate at the 0.1\% level by spectroscopy of the various modes and estimate the tension in the range of 0.8 - 0.9 GPa for the various samples studied. 

The mechanical quality factors for various modes are measured through ringdown measurements of the membrane oscillation. For this, the membrane is piezo-actuated at the various membrane resonances up to amplitudes of around 200 pm for durations of 10 ms before switching off the drive. The amplitude of membrane oscillation is then monitored through a lock-in amplifier to measure the $(1/e)$ decay time $\tau$ of the oscillation amplitude. The quality factor is then estimated as $Q_{jk} = \omega_{jk} \tau/2$ where $\omega_{jk}$ is the eigenfrequency of the mode under study. 

Various checks of systematic effects on these measurements were made to ensure the validity of our interpretation. These include the negligible effect of radiation pressure or photothermal heating due to the laser field on the mechanical motion, the linearity of the drive as well as the negligible influence of viscous damping at our operating background pressure of $p \sim 2 \times 10^{-7}$ Torr (see Supplementary Information for details). Changes in the peak amplitude by up to a factor of five in either direction does not change the measured value of the quality factor indicating that we are operating far from any intrinsic nonlinearities of the mechanical resonator. For the typical amplitudes of mechanical motion during the ring-down measurements, self-stiffening nonlinearities were measured to be below the 1 ppm level. Finally, the mechanical linewidth inferred from thermal Brownian motion is consistent with that derived from the ringdown measurements. 

\begin{figure}
\includegraphics[width=2.75in]{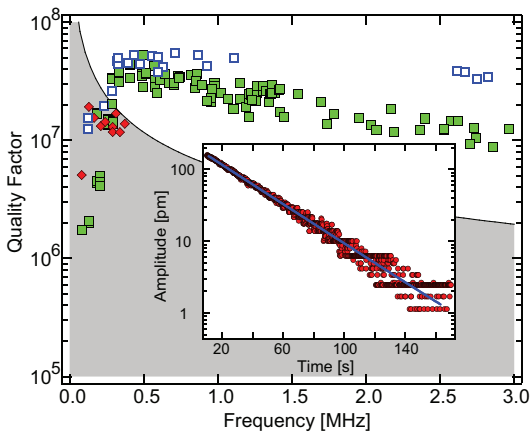}
\caption{Peak mechanical quality factors of a $L=5$ mm, $h = 100$ nm SiN membrane versus frequency ($\blacksquare$). The solid line corresponds to $f \times Q = k_B/h \times $ 300K. For low mode frequencies ($\nu_{jk} < 300$ kHz) , the quality factors can be improved by an order of magnitude ($\blacklozenge$) simply by reducing the contact region between the substrate and the in-vacuum mount. The weak frequency dependence of the measured $Q$s at high frequencies is further reduced for $h=30$ nm ($\Box$), (see text for discussion). 
Inset: Characteristic mechanical ringdown of the (6,6) mode at $\nu_{66} = 678$ kHz.}
\label{Fig:Qs}
\end{figure}

Fig. 1 shows the typical trend observed for the peak quality factors of various mechanical modes exhibited by the membrane. We typically extend our measurements up to mode indices $(j^2 + k^2)^{1/2} \sim 40$. For higher mode indices, the rapidly increasing density of modes renders it challenging to accurately resolve individual modes of the resonator. More importantly, we also observe substantial intermodal coupling between proximal eigenmodes which complicates the interpretation of the measured quality factors. 

For our measurements, we distinguish between two regimes of behavior : (i) For mode indices $(j^2 + k^2)^{1/2} < 4$, the measured $Q$s exhibit highly non-monotonic behavior with increasing frequency. We also observe a large variation and a sensitive dependence of the quality factors on the clamping mechanism as well as the geometry of the modal structure. Our observations in this regime are quantitatively consistent with the dominant loss mechanism being anchor losses from the membrane into the supporting mount \cite{wilsonrae2008}. The sensitive dependence of the $Q$s on the clamping mechanism can be greatly reduced ($\blacklozenge$,Fig.\,1) by ensuring minimal contact between the supporting silicon wafer and the in-vacuum mount, reinforcing the above interpretation. (ii) For mode indices $(j^2 + k^2)^{1/2} \gtrsim 4$, the peak $Q$s reach a plateau around $5 \times 10^7$. The peak quality factors in this regime are a weak function of the resonant frequencies with a scaling estimated as $Q_{jk} \sim \nu_{jk}^{-(0.7 \pm 0.15)}$ for the 100 nm membranes. This scaling becomes much weaker as the thickness of the membrane is reduced ($\Box$, Fig.1), resulting in $Q$s that are almost independent of frequency. In this latter regime, the measured $Q$s are more robust to variations in the clamp making it less obvious that direct anchor loss is the dominant loss mechanism. 

\begin{figure}
\includegraphics[width=2.5in]{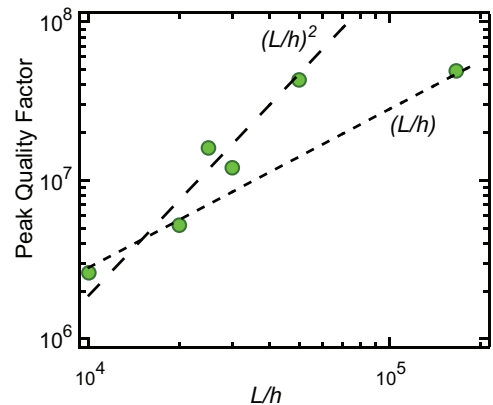}
\caption{Peak mechanical quality factors versus film geometry parametrized by the ratio of membrane width ($L$) to membrane thickness ($h$). For $L/h < 10^5$, we observe a scaling consistent with $Q \sim (L/h)^2$. A linear scaling is also shown for reference.}
\label{Fig:Peak_Q}
\end{figure}

We have measured the peak $Q$s in the plateau regime for a range of membrane geometries ranging in width from $L = 0.5 - 5$ mm and in thickness from $h=30 - 200$ nm (Fig.2). For $L/h < 10^5$, we find the following scaling relations $Q \sim (L/h)^2$ and  $f \times Q \sim (L/h)$. The noticeable discrepancy in this scaling for the thinnest membranes is, at present, unexplained \cite{XPSfootnote}. 

We have performed a series of experiments to elucidate the limiting damping process of the mechanical excitations in the plateau regime. In the context of our system, the possible range of mechanisms include thermoelastic damping (TED) \cite{lifshitz2000, photiadis2002, norris2005}, damping due to localized defects on the surface or within the bulk of the resonator \cite{seoanez2008, unter2010}, anharmonic processes such as Akhiezer damping \cite{akhiezer1939} and anchor (radiation) losses from the resonator to the supporting structure \cite{park2004,wilsonrae2008, jockel2011,cole2011}. Based on models and experimental measurements described below, we find quantitative evidence that our current quality factors are limited by anchor losses from the membrane into the substrate. 

\begin{figure}[h]
\includegraphics[width=3.5in]{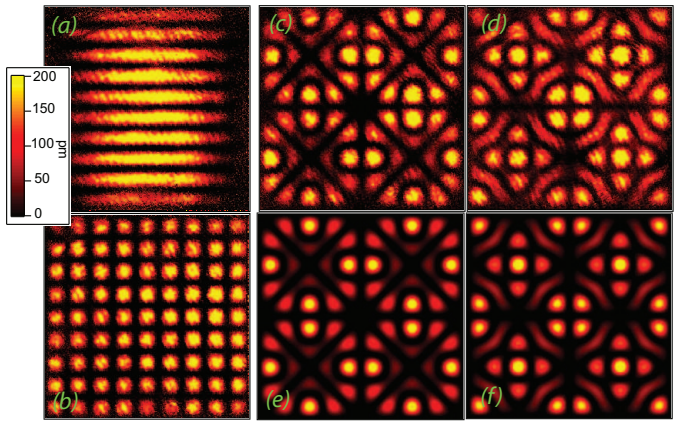}
\caption{Interferometric imaging of the mechanical modes : In situ images of the (a) (1,10) mode and (b) (9,9) mode (Color scale for displacement shown). Substrate-induced coupling between proximal asymmetric eigenmodes result in hybridization into more symmetric structures. (c),(d) show the modal structures corresponding to $\phi_{10,6} \pm \phi_{6,10}$ hybridized modes. These can be compared to the calculated mode profiles for this hybridization (e,f). }
\label{Fig:mag_growth}
\end{figure}

Firstly, we have developed a model of thermoelastic dissipation in our membrane resonators that takes into account their large intrinsic tensile stress of $\sim1$GPa. Within this model, the mechanical motion of the membrane couples to a local temperature field associated with microscopic changes in the volume of the resonator. This results in local irreversible heat flows and dissipation (see Supplementary Information for more details). Our model is built upon the formalism introduced in \cite{lifshitz2000, nayfeh2004} and accurately reproduces the eigenfrequencies of our resonators for the entire range of geometries studied. Importantly, for established material parameters of stoichiometric SiN, our model predicts a room temperature limit of $Q_{TED} \sim 10^{12}$ for the frequency ranges of our study. Further, our model also predicts that the quality factor scales as $Q \sim 1/\nu_{jk}^2$ in the frequency regime $\nu_{jk}/\nu_{11} > 10$ in distinct contrast to the weak dependence experimentally observed (Fig.1). Based on these model predictions, we discount TED in the membrane as a limiting influence of our observed quality factors. 

Another source of dissipation is the coupling between the mechanical motion and intrinsic, localized defects within the membrane. While the microscopic origins of these defects remain unknown, these losses are typically modeled as being due to two level systems (TLS) whose energy splitting is modulated by the oscillating strain field \cite{jackle1972}. The subsequent re-equilibration of these two level systems results in attenuation of the mechanical energy. At the elevated temperatures of our experiments, these two-level systems can be regarded as being thermally activated over a wide range of energy scales. For the mechanical frequencies in this work, the TLS model \cite{tielburger1992} predicts a scaling of $Q_{jk} \sim 1/\nu_{jk}$ and a dissipation that scales very weakly with the modal structures, dimensions of the resonator and details of the support structure. These predictions are inconsistent with our observations (see Fig. 2,4). 

The quality factors for a given sample remain stable within 10\% of the measured values even after exposure to air for several days. We have also annealed the membranes at temperatures up to 650$^\circ$C under vacuum to reduce surface contamination without a significant change in quality factors. These observations rule out surface-induced losses as a contributing influence.

In addition to the above observations that rule out specific intrinsic processes, a more mechanism-agnostic argument can be formulated for a large class of intrinsic dissipation mechanisms. For any such process that couples mechanical motion to a source of dissipation, the leading symmetry-allowed term in the equation of motion must be proportional to the local curvature of the displacement field (see, for example \cite{yu2012}). Thus, the measured quality factors for the various modes should correlate with the local modal curvature or higher powers thereof. In order to better quantify this reasoning, we have developed an interferometric imaging technique capable of spatially resolving the modal structure of the resonator \cite{imaging2013}. These images (Fig.3) yield a wealth of spatial information complementing our spectroscopic measurements. In the plateau regime, we observe that the quality factors can vary by almost two orders of magnitude for a corresponding variation in the integrated curvature of less than 20\% pointing to an extremely weak correlation between the two quantities.


We also note that we have measured $Q$s up to $2.7 \times 10^7$ in low stress SiN membranes ($\sigma = 0.25$ GPa) of similar geometry, i.e. within a factor of two of those measured in the high stress membranes. Further, we observe a substantial influence of the substrate on the membrane modes (see Fig.3) with nominally degenerate eigenmodes hybridizing into more symmetric structures. At larger drive amplitudes than those used in this study, we observe multimode bistability as a result of such coupling to the substrate \cite{bistab}. Based on the preceding arguments, we conclude that there is little or no influence of intrinsic material processes on our measured $Q$s. 

Finally, we discuss the role of radiation loss from the membrane into the supporting substrate. This loss of mechanical energy arising from the coupling to the external supporting substrate represents one of the fundamental restrictions to a high-$Q$ device that is only weakly dependent on the material parameters of the resonator. Accordingly, various treatments and models of this radiation loss \cite{park2004, photiadis2004, judge2007} have been developed to address design methodologies that can alleviate this loss. A particularly intuitive picture of `phonon tunneling' \cite{wilsonrae2008, wilsonrae2011} has recently emerged, wherein the resonator can be regarded as a phononic cavity coupled to the external substrate through a weak coupling parameter. 


Guided by our $Q$-factor measurements and modal images, we have extended this model to the higher mode indices of the plateau regime and obtain excellent agreement (see Supplementary Information for details). In order to decouple the geometrical dependence of the quality factor from other frequency-dependent factors, we compare the model to $Q$s measured over a narrow range of frequency corresponding to an arc of radius $\sqrt{j^2 + k^2} \sim 12$ in the $(j,k)$ space. In particular, we see that the substrate-induced mode coupling and ensuing hybridization (see Fig. 3) suppresses the large $Q$ variation between modes of even and odd parity that is seen at lower frequencies. Instead, we observe a more gradual variation with lower $Q$s measured for  modes with either $j \ll k$ or $j \gg k$, and higher $Q$s measured for $j \sim k$ (see Fig. 4). The close agreement between our observations and the tunneling model (suitably modified to include interference effects arising from substrate-induced hybridization), further clarifies the dominant role of anchor losses in determining our peak quality factors. 

\begin{figure}
\includegraphics[width=3.25in]{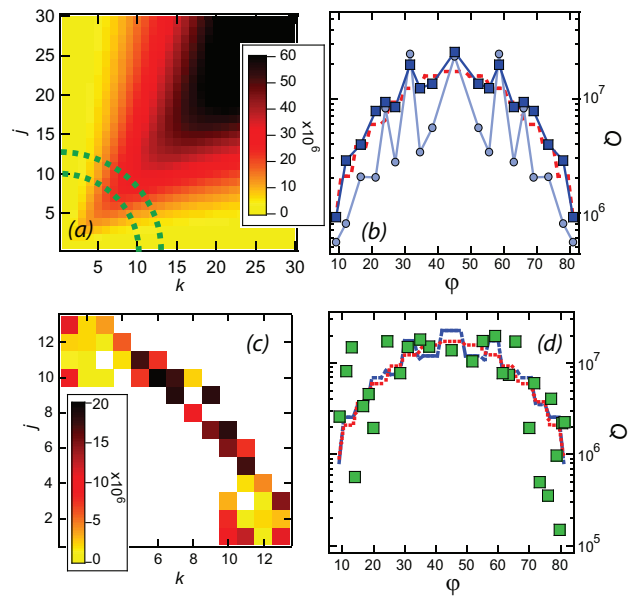}
\caption{(a) Predicted quality factors versus mode indices based on asymptotic limits of our anchor loss model (see Supplementary Information for details), (b) Predicted quality factors versus $\phi \equiv \arctan(j/k)$ for mode indices $(j^2 + k^2)^{1/2} = 12$ with (without) substrate-mediated hybridization are shown as $\blacksquare$ ($\bullet$). Also shown is the asymptotic expression for the quality factors from our model (dashed line). (c) Measured $Q$s for mode indices indicated by the green arc in (a). (d) Comparison between the measured $Q$s in this arc and our predictions from (b). }
\label{Fig:mag_growth}
\end{figure}

Our modified treatment of the phonon tunneling model accurately captures the increased robustness and enhanced quality factors arising from the interference between nominally degenerate resonator modes. One can extend this logic to the complementary scenario, i.e. where interference between distinct substrate modes lead to a similar enhancement of the quality factors. The extreme limit of this latter scenario is when the substrate exhibits an acoustic bandgap in the vicinity of the relevant resonator modes. This should lead to a vastly enhanced quality factor and $f \times Q$ product that is limited, in principle, only by the material properties of the membrane. 

In summary, we demonstrate stoichiometric SiN membrane resonators with ultrahigh quality factors up to $5 \times 10^7$ and $f \times Q \sim 10^{14}$ Hz that is, to our knowledge, the largest reported in a room temperature mechanical system. We have developed models of our system that identify anchor loss as the dominant decay mechanism that limits the current quality factors. Remarkably, the material properties of stoichiometric Silicon nitride do not limit the performance of these membrane resonators even at such high quality factors. The true intrinsic material limitations of stoichiometric SiN on the quality factor and $f \times Q$ products of our resonators remain a topic for further study. 

Our current system realizes an enabling platform for the optomechanical cooling and quantum control of a mesoscopic resonator at room temperature. Further, the low mechanical frequencies and ultrahigh quality factors demonstrated in this work are ideally suited to various schemes to interface atomic gases or solid state spin systems to the mechanical degree of freedom , thereby realizing hybrid quantum devices for sensor and transduction applications as well as for fundamental studies \cite{steinke2011}. 

We acknowledge S. Vengallatore for valuable discussions, A. K. Bhat for experimental assistance and E. J. Mueller for a critical reading of the manuscript. This work was supported by the DARPA QuASAR program through a grant from the ARO and the Cornell Center for Materials Research with funding from the NSF MRSEC program (DMR-1120296). M. V. acknowledges support from the Alfred P. Sloan Foundation. 

\bibliography{SiNbib, SiNnotes}

\begin{thebibliography}{36}
\bibitem [{XPS()}]{XPSfootnote}%
  \BibitemOpen
  \href@noop {} {}\bibinfo {note} {Through X-ray photoemission spectroscopy, we
  have verified that the material composition of the membranes of various
  thicknesses are identical. High resolution images of the 30 nm membranes
  reveal no visible sign of damage. Further, the measured eigenfreqencies are
  consistently within 2\% of those measured for the thicker membranes of the
  same width indicating that the tensile stress is of similar
  magnitude.}\BibitemShut {Stop}%
\end{thebibliography}

\section*{Supplementary Information}
\subsection*{Sample preparation}
The membrane resonators are placed on an aluminum ring approximately 1.5 cm in diameter and glued to this ring along the edge of the supporting silicon substrate using an epoxy such as Torr-seal. We take care to use very small amounts of the epoxy and have found that excessive use of glue along the edges can reduce the measured quality factor. For most of the studies described in this work, the silicon substrate makes contact with the Aluminum ring only along two corners. The aluminum ring is then glued to the front of a ring-piezo actuator and placed within an UHV chamber for our ringdown studies. 

\subsection*{Viscous damping limits on Quality factor}
For the operating background pressures in our studies, the limits on the mechanical quality factor due to viscous damping from background gas can be estimated as \cite{bao2002}
\begin{equation}
Q_{jk}(p) = \left( \frac{\pi}{2} \right)^{3/2} \rho h \nu_{jk} \sqrt{\frac{k_B T}{m_g}} \frac{1}{p}
\end{equation}
where $p$ is the background pressure and $m$ is the molecular mass of the gas molecules in the background. For our operating base pressure of $p \sim 2 \times 10^{-7}$, this limit evaluates to around $Q \sim 10^9$ for modal frequencies around $\omega = 2 \pi \times 500$ kHz.

\subsection*{Measurement of optical properties of the SiN membranes}
The imaginary part of the refractive index was estimated by placing the membrane resonators within a high finesse Fabry-Perot cavity and measuring the corresponding decrease in the finesse due to absorptive losses in the membrane. For these measurements, we made use of a Fabry-Perot cavity characterized by a finesse of $115,000$ corresponding to a cavity linewidth of $\kappa = 2 \pi \times 70$ kHz. A stoichiometric SiN membrane of thickness $h = 200$ nm was placed within this cavity. This choice of membrane thickness resulted in a significant modification of the bare cavity parameters due to the optical losses within the membrane. Perhaps more importantly, the membrane reflectivity for this particular thickness is close to zero at our operating wavelength of $\lambda = 795$ nm, (i.e. $2 \pi \mathcal{R}(n) h/\lambda \sim \pi \Rightarrow r \sim 0$). This ensures a minimal dependence of the modified cavity finesse on the alignment and position of the membrane with respect to the intracavity field. 

Over a range of positions of the membrane within the Fabry-Perot cavity, we measured a modified finesse of $70,000$ corresponding to a cavity linewidth of $\kappa' = 2 \pi \times 105$ kHz.  Under the assumption that this modification is entirely due to absorptive losses within the membrane, our measurements are consistent with $\mathcal{I}(n) < 9 \times 10^{-6}$. 

\subsection*{Parity dependence of Quality factors for low mode indices}
For low mode indices, we observe that the measured quality factors can show large variations depending on the parity of the modes (see Fig.\,S1). Also, in contrast to our observations of the modes in the plateau regime, we find that the modal structures for these low mode indices are negligibly influenced by the substrate. Thus, the phonon tunneling model as described in \cite{wilsonrae2008, wilsonrae2011} accurately predicts our measured quality factors in this regime. In contrast, as the mode indices grow larger, interferences between degenerate eigenmodes arising from coupling to the substrate result in a suppression of the large parity-dependent variation of the quality factors and a more robust occurence of a large number of high-$Q$ modes. 

\begin{suppfig}[t]
\includegraphics[width=2.5in]{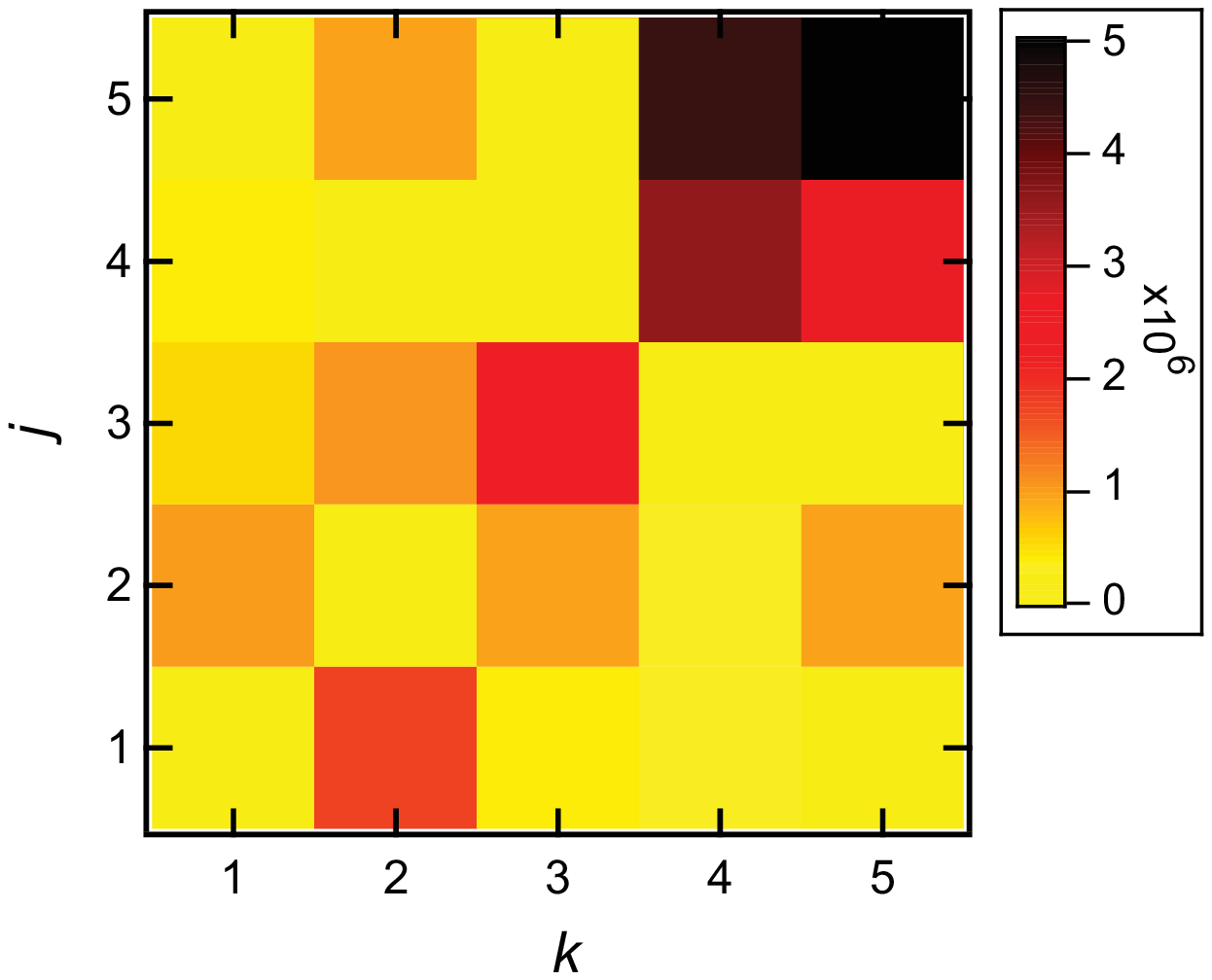}
\caption{Parity dependence of the measured quality factors for low mode indices ($j, k \leq 5$).}
\label{Fig:mag_growth}
\end{suppfig}

\subsection*{Themoelastic damping limits}
In this section, we describe our model for thermoelastic damping in the high stress membranes. Our model follows the formalism outlined in \cite{lifshitz2000, nayfeh2004}. The equation of motion for a membrane under tension is
\begin{equation}
D\left( \frac{\partial^4 w}{\partial x^4} + \frac{\partial^4 w}{\partial y^4} \right) - \sigma h \left( \frac{\partial^2 w}{\partial x^2} + \frac{\partial^2 w}{\partial y^2} \right) = -\rho h \frac{\partial^2 w}{\partial t^2}
\end{equation}
where $w(x,y)$ is the displacement of the membrane in the $\hat{z}$ direction and is assumed to be much smaller than the membrane thickness $h \sim 100$ nm. The lateral dimension of the membrane is given by the width $L$, the membrane tension is denoted by $\sigma$ and $\rho = 2.7$ g/cm$^3$ is the density of stoichiometric silicon nitride. The flexural rigidity $D$ is related to the Young's modulus ($E$), the thickness $h$ and the Poisson's ration $\nu_P$ through the expression
\begin{equation}
D = \frac{E h^3}{1 - \nu_P^2}
\end{equation}

In the presence of thermoelastic damping, the local temperature variations within the membrane and the resulting thermal expansion lead to additional stresses. The equation of motion in the presence of thermal strain is given by \cite{nayfeh2004}
\begin{equation}
D \nabla^4 w - \sigma h \nabla^2 w + \rho h \frac{\partial^2 w}{\partial t^2} = -\nabla^2 M^T - N^T \nabla^2 w 
\end{equation}
where the thermal axial force and the thermal bending moment are respectively given by
\begin{eqnarray}
N^T &=& \frac{E \alpha}{1 - \nu_P} \int \theta dz \\
M^T &=& \frac{E \alpha}{1 - \nu_P} \int z \theta dz
\end{eqnarray}
where $\theta(x,y,z,t) = T(x,y,z,t) - T_0$ is the local deviation in temperature, $T_0$ is the equilibrium temperature and $\alpha$ is the coefficient of thermal expansion. The local temperature field itself satisfies the diffusion equation given by
\begin{equation}
\kappa \nabla^2 \theta = \rho C_P \frac{\partial \theta}{\partial t} - \frac{E \alpha T_0}{1 - \nu_P} \frac{\partial}{\partial t} (z \nabla^2 w)
\end{equation}
 where $\kappa$ is the thermal conductivity and $C_P$ is the specific heat. The coupled equations (1),(6) can be solved for the normal modes of vibration in the presence of thermoelastic damping. The quality factor can then be extracted from the corresponding eigenfrequencies as 
 \begin{equation}
 Q^{-1}(\omega) = 2 \frac{\mathcal{I}(\omega)}{\mathcal{R}(\omega)}
 \end{equation}
 
 We solve the diffusion equation for the local temperature field with boundary conditions corresponding to no heat flow at the top and bottom surfaces. This yields
 \begin{equation}
 \theta(x,y) = \frac{E \alpha T_0}{(1-\nu_P) \rho C_P} \nabla^2 w(x,y) \left( z - \frac{\sin(\tilde{k} z)}{\tilde{k} \cos(\tilde{k} h/2)} \right)
 \end{equation}
 where $\tilde{k} = (1 - i) \sqrt{\frac{\omega \rho C_P}{2 \kappa}}$. Substituting this solution into the expressions for the thermal strain and bending moment results in a modified equation for the membrane motion
 \begin{equation}
 (D + D_t) \nabla^4 w_0 - \sigma h \nabla^2 w_0 = \rho h \omega^2 w_0
 \end{equation}
 where we have written $w(x,y,t) = w_0(x,y) e^{i \omega t}$ and 
 \begin{equation}
 D_t = \frac{E^2 \alpha^2 T_0}{(1 - \nu_P)^2 \rho C_P} \left( \frac{h^3}{12} + \frac{h}{\tilde{k}^2} - \frac{2 \tan (\tilde{k} h/2)}{\tilde{k}^3} \right)
 \end{equation}
is the correction due to thermoelasticity. The damping of the membrane is encapsulated in the non-zero imaginary part of this term. Solving the equation of motion for the normal modes yields the eigenfrequencies
\begin{equation}
\omega_{mn} = \sqrt{\frac{\sigma \pi^2}{\rho L^2} (m^2 + n^2) + \frac{(D + D_t)}{\rho h L^4} (m^2 + n^2)^2 \pi^4}
\end{equation}
In the case of large tensile stress, i.e. in the regime where $\eta \equiv \frac{D}{\sigma h L^2} \ll 1$, the expression for the eigenfrequencies reduces to 
\begin{equation}
\omega_{mn} = \omega^0_{mn} \left( 1 + \frac{\pi^2 (D + D_t)}{2 \sigma h L^2} (m^2 + n^2) \right)
\end{equation}
where $\omega^0_{mn} = 2 \pi \sqrt{\frac{\sigma}{4 \rho L^2} (m^2 + n^2)}$. The quality factor is then given by
\begin{equation}
Q^{-1}_{mn} \approx \frac{\pi^2 (m^2 + n^2)}{\sigma h L^2} \mathcal{I}[D_t]
\end{equation}
where
\begin{eqnarray}
\mathcal{I}[D_t] &=& \frac{E^2 \alpha^2 T_0 h^3}{12 (1 - \nu_P)^2 \rho C_P} g(\xi) \\
g(\xi) &=& \frac{6}{\xi^3} \frac{\sinh \xi + \sin \xi}{\cosh \xi + \cos \xi} - \frac{6}{\xi^2}
\end{eqnarray}
and $\xi = h (\omega \rho C_P/2 \kappa)^{1/2}$. Written in terms of the undamped natural eigenfrequency $\omega^0_{mn}$ of the resonator, the thermoelastic damping limit on the quality factor finally reduces to 
\begin{equation}
Q(\omega_{mn}) = 12 \frac{(1 - \nu_P)^2 C_P \sigma^2}{(\omega^0_{mn})^2 E^2 \alpha^2 T_0 h^2} \frac{1}{g(\xi)}
\end{equation}
Using bulk material parameters for stoichiometric Silicon nitride and for a typical membrane thickness of $h = 100$ nm, this limit evaluates to $Q_{TED} \sim 10^{12}$ for modal frequencies around $\omega = 2 \pi \times 500$ kHz. This estimate is consistent with that derived by Zwickl et al \cite{zwickl2011}, but significantly higher than the estimate in Wilson et al \cite{wilson2009} which neglects the large intrinsic stress of the SiN membranes. 

\subsection*{Modal image analysis and estimate of local curvature}
Here, we describe our method of estimating the local curvature from the interferometric images of the various resonator modes. To image a particular mode, the membrane is driven on resonance to an amplitude of around 200 pm. An imaging beam of 3.5 mm waist and a power of 10 $\mu$W is incident on the membrane. The DC component of the reflected light is filtered by means of a 200 $\mu$m spot and the resulting dark field image is captured on a CCD camera with typical exposure times of 1 ms. The amplitude of motion is calibrated with a Michelson interferometer with a smaller spot size focused at the center of the membrane. In addition to each modal image ($I(x,y)$), images are also obtained of the background, i.e. of the undriven membrane ($I_{bkgd}$) and the laser beam profile ($I_{beam}$). 

The image proportional to the modulus of the displacement is given by
\begin{equation}
|\tilde{w}(x,y)| = \sqrt{\frac{I(x,y) - I_{bkgd}(x,y)}{I_{beam}(x,y)}}
\end{equation}
The local curvature $C(x,y)$ is obtained by taking the finite differences from the $266 \times 266$ pixel image according to the expression
\begin{eqnarray}
C(x,y) &\rightarrow& C[i,j] = \frac{\tilde{w}[i+d,j] + \tilde{w}[i-d,j] - 2 \tilde{w}[i,j]}{d^2} \nonumber \\
 &+& \frac{\tilde{w}[i,j+d] + \tilde{w}[i,j-d] - 2 \tilde{w}[i,j]}{d^2}
\end{eqnarray}
where an optimal value of $d = 5$ pix was used as a compromise between signal to noise and spatial resolution. 

As noted earlier, an intrinsic loss mechanism would contribute terms in the equation of motion that, to leading order, are proportional to the local curvature of the displacement field. Accordingly, the energy loss in the membrane due to such mechanisms is given by
\begin{equation}
\Delta E \propto \int dx dy \left( \frac{\partial^2 w}{\partial x^2} + \frac{\partial^2 w}{\partial y^2}  \right)^2
\end{equation} 
while the stored energy in the membrane can be written as $E = \frac{1}{2} \rho h \omega_{jk}^2 \int dx dy \, w(x,y)^2$. The quality factor associated with such an intrinsic loss mechanism is then given by
\begin{eqnarray}
Q &=& \frac{2 \pi E}{\Delta E} \propto \frac{\int dx dy \, w(x,y)^2}{\int dx dy \, \left( \frac{\partial^2 w}{\partial x^2} + \frac{\partial^2 w}{\partial y^2} \right)^2} \nonumber \\
 &\rightarrow& \frac{\sum_{i,j} \tilde{w}[i,j]^2}{\sum_{i,j} C[i,j]^2}
\end{eqnarray}
By construction, the above quantity does not depend on the magnitude of the membrane displacement. 

\subsection*{Anchor loss model for hybridized resonator modes}
In this section, we outline our analysis of the anchor loss model that takes into account the interference effects due to substrate-mediated hybridization. The `phonon tunneling' model introduced in \cite{wilsonrae2008} permits a general expression for the clamping-limited quality factor $Q$ in terms of the modal overlap between the resonator and the free substrate
\begin{equation}
Q^{-1}(\omega_R) = \frac{\pi}{2 \rho_s \rho_R \omega_R^3} \int_q \left | \int_S d \vec{S} \cdot \breve{{\pmb \sigma}}_R \cdot \vec{u}^0_q \right |^2 \times \delta(\omega_R - \omega(q))
\end{equation}
where $\rho_s (\rho_R)$ are the mass densities of the substrate (resonator), $\omega_R$ is the resonant frequency, $\breve{{\pmb \sigma}}_R$ is the stress field of the resonator and $\vec{u}^0_q$ is the displacement field of the free substrate modes. 

The large intrinsic stress of the stoichiometric SiN and the extremely small thickness of the membrane allow us to relate the stress field at the periphery to the slope of the resonator mode evaluated at the clamp \cite{wilsonrae2011}, i.e. 
\begin{equation}
\int \hat{z} \cdot \breve{{\pmb \sigma}}_R \cdot \hat{x} dz = \sigma \sqrt{h} \frac{\partial \phi_R}{\partial x}
\end{equation}
and a similar relation for $\hat{y}$. Here, $\sigma$ is the intrinsic stress of the silicon nitride membrane and $\phi_R(x,y)$ is the resonator eigenmode. For the square membranes, this approximation results in
\begin{eqnarray}
\mathcal{S} \equiv \int_S d\vec{S} \cdot \breve{{\pmb \sigma}} \cdot \vec{u}_q^0 &=& 4 \sigma \sqrt{h} \int  dx \,  \vec{u}(x, y=L/2) \frac{\partial \phi_R}{\partial y} \rvert_{y=L/2} \nonumber \\
 &+& 4 \sigma \sqrt{h} \int  dy \,  \vec{u}(x=L/2, y) \frac{\partial \phi_R}{\partial x} \rvert_{x=L/2} \nonumber
\end{eqnarray}

\begin{suppfig}[t]
\includegraphics[width=3.25in]{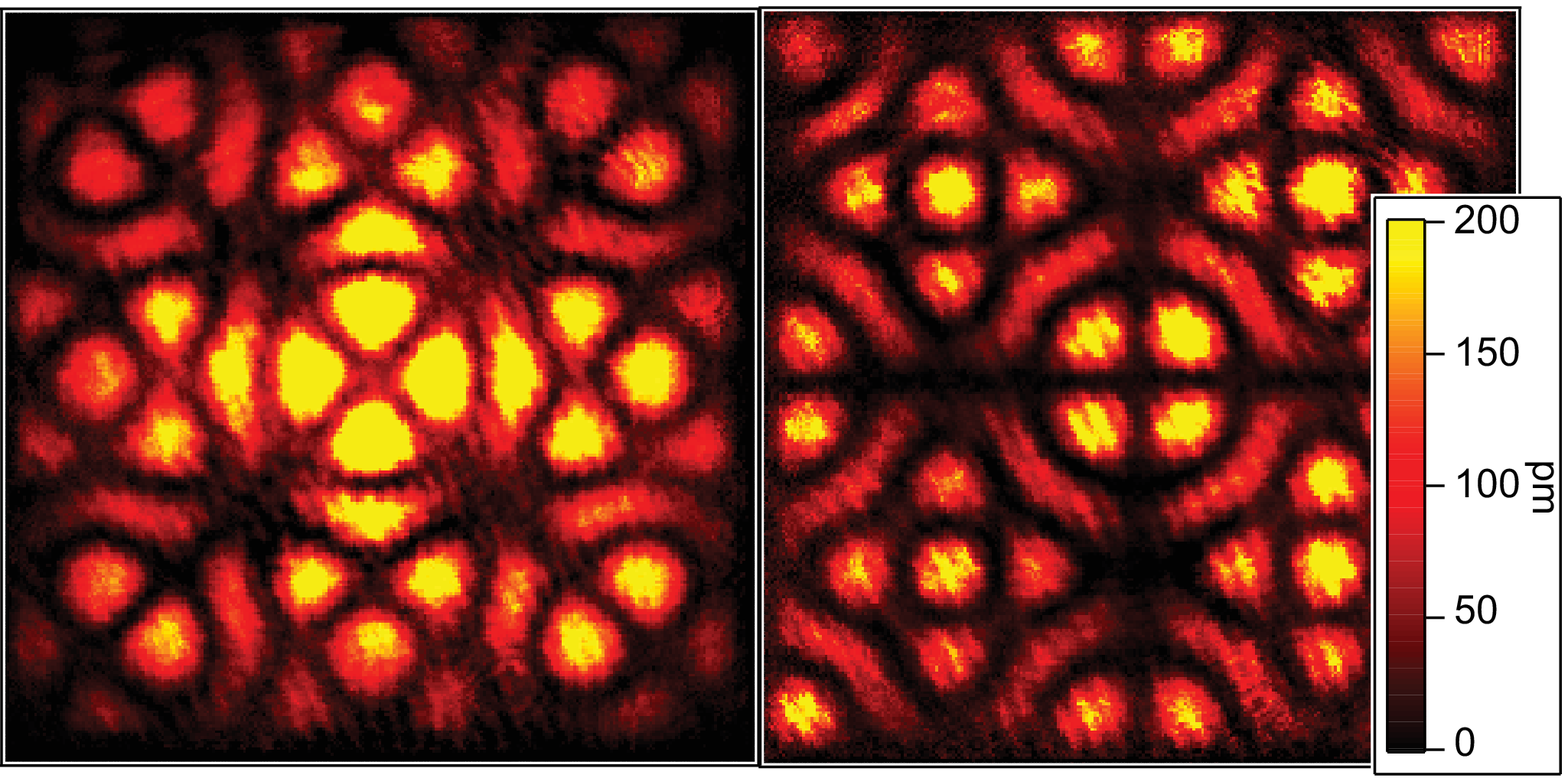}
\caption{Interferometric images of the modal structures resulting from substrate-mediated coupling. Left: Hybridized mode corresponding to $\phi_{3,11} - \phi_{11,3}$, Right: Hybridized mode corresponding to $\phi_{10,6} + \phi_{6,10}$.}
\label{Fig:mag_growth}
\end{suppfig}

Symmetry considerations imply that contributions to the surface integral only arise from substrate and resonator modes that share the same symmetry, i.e. resonator modes with even parity only couple to substrate modes with even parity etc. In the case of the `single-mode' calculation, this consideration results in distinguishing among three distinct cases, (i) Symmetric modes ($j, k$ are odd), (ii) Antisymmetric modes ($j,k$ are even) and (iii) Mixed symmetry modes ($j$ even, $k$ odd or vice versa). Accordingly, this results in separate expressions of the anchor-limited quality factors with different angular dependences. 

In contrast, the hybridized modes exhibit highly symmetric modal structures that is less dependent on the parity of the mode indices (see, for example, Fig.\,S2). This results in a suppression of the radiated substrate modes and a concomitant enhancement of the quality factor for a larger class of resonator modes. 

The full calculation that takes this enhanced symmetry into account, while laborious, is fairly straightforward and rapidly convergent in the limit $j, k > \eta_\gamma \sqrt{j^2 + k^2}$. The results of this calculation for modes $\sqrt{j^2 + k^2} = 12$ are shown ($\bullet$) in Fig.\,4(b). Also shown alongside is the calculation that does not take into account any modal hybridization. The contrast between the two predictions is especially stark for modes of low symmetry. The `single-mode' calculation predicts a large variation of $Q$ with alternating parity while the modified calculation suppresses this variation resulting in uniformly higher $Q$s. 

For large mode indices $j, k \gg 1$ with $j, k > \eta_\gamma \sqrt{j^2 + k^2}$, the suppressed parity dependence can be well approximated by the asymptotic expression
\begin{equation}
Q_{kj, asympt}^{-1} \approx \frac{16 \pi j^2 k^2}{\sqrt{j^2 + k^2}} \left[ \frac{1}{j^4} + \frac{1}{k^4} \right] \frac{\rho_R h}{\rho_s L}  \sum_\gamma \eta^3_\gamma |u_\gamma^{(0)}|^2 
\end{equation}
where the summation is now carried out over the various modes (surface acoustic, transverse, longitudinal and flexural waves) supported by the substrate. Here, $\eta_\gamma = c_R/c_\gamma \ll 1$ is the ratio of phase velocities in the membrane and the substrate. Note that, in this limit, the summation evaluates to a geometry-independent constant and the entire modal dependence is contained in the pre-factor. For comparison, this approximate expression is plotted alongside the exact calculation in Fig.\,4(b). 


Intuitively, hybridization between degenerate resonator modes results in modal structures that are more symmetric than either of the constituent modes (see Fig.\,S2). This enhanced symmetry results in two obvious trends that are captured in Fig.\,4(b), i.e. (i) An increased quality factor due to destructive interference of substrate modes radiated from the evenly spaced antinodal segments around the clamp and (ii) A suppressed dependence of $Q$ on the parity of the resonator mode. 

\end{document}